# Surface passivation of FAPbI$_3$-rich perovskite with caesium iodide outperforms bulk incorporation.

Thomas P. Baumeler[1,+], Essa A. Alharbi[1,2,+,*], George Kakavelakis[1], George C. Fish[3], Mubarak T. Aldosari[2], Miqad S. Albishi[2], Lukas Pfeifer[1], Brian I. Carlsen[1], Jun-Ho Yum[4], Abdullah S. Alharbi[2], Mounir D. Mensi[5], Jing Gao[1], Felix T. Eickemeyer[1], Kevin Sivula[4], Jacques-Edouard Moser[3], Shaik M. Zakeeruddin[1] and Michael Grätzel[1,*]

[1] Laboratory of Photonics and Interfaces, Institute of Chemical Sciences and Engineering, School of Basic Sciences, Ecole Polytechnique Fédérale de Lausanne, CH-1015, Lausanne, Switzerland

[2] Microelectronics and Semiconductors Institute, King Abdulaziz City for Science and Technology (KACST), Riyadh 11442, Saudi Arabia

[3] Photochemical Dynamics Group, Institute of Chemical Sciences and Engineering, École Polytechnique Fédérale de Lausanne, Lausanne 1015, Switzerland

[4] Laboratory of Molecular Engineering of Optoelectronic Nanomaterials, Institute of Chemical Sciences and Engineering, École Polytechnique Fédérale de Lausanne (EPFL), Lausanne 1015, Switzerland

[5] Institute of Chemical Sciences and Engineering, École Polytechnique Fédérale de Lausanne, Valais, CH-1951 Sion, Switzerland

[+] These authors contributed equally to this work

[*] Correspondence to Michael Grätzel (michael.graetzel@epfl.ch), and Essa A. Alharbi (ealharbi@kacst.edu.sa).

## Abstract

Metal halide perovskites (MHPs) have shown an incredible rise in efficiency, reaching as high as 25.7%, which now competes with traditional photovoltaic technologies. Herein, we excluded CsX and RbX (X=I$^-$, Br$^-$, Cl$^-$), the most commonly used cations to stabilize α-FAPbI$_3$, from the bulk of perovskite thin films and applied them on the surface, as passivation agents. Extensive device optimization led to a power conversion efficiency (PCE) of 24.1% with a high fill factor (FF) of 82.2% upon passivation with CsI. We investigated in-depth the effect of CsI passivation on structural and optoelectronic properties using X-ray diffraction (XRD), angle resolved X-ray photoelectron spectroscopy (ARXPS), Kelvin Probe Force (KPFM) microscopy, time-resolved photoluminescence (TRPL), photoluminescence quantum yield (PLQY) and electroabsorption spectroscopy (TREAS). Furthermore, passivated devices exhibit enhanced operational stability, with optimized passivation with CsI leading to a retention of ~90% of initial PCE under 1 Sun illumination with maximum power point tracking for 600 h.

## Introduction

Metal halide perovskite (MHPs) solar cells are attracting tremendous attention from both industry and academia as one of the most promising emerging photovoltaic technologies.[1–4] From an initial power conversion efficiency (PCE) of 3.8% in 2009[5], single-junction perovskite solar cells (PSCs) rapidly evolved to reach a current record PCE value of 25.7%.[6] MHPs are a class of material with the general formula ABX$_3$; A stands for a monovalent cation, whether organic (HC(NH$_2$)$_2$$^+$ (FA$^+$), CH$_3$NH$_3$$^+$ (MA$^+$)) or inorganic (Cs$^+$, Rb$^+$), B for a divalent metal cation (Pb$^{2+}$ or Sn$^{2+}$) and X for a monovalent halide anion (I$^-$, Br$^-$, Cl$^-$). Among a wide range of different perovskite compositions, the cubic α-phase of formamidinium lead iodide (α-FAPbI$_3$) represents the best candidate for highly efficient single-junction solar cells fabrication due to its narrow bandgap of 1.45 to 1.53 eV[7–9], close to the optimum value given by the Shockley-Queisser limit, and higher thermal stability relative to MAPbI$_3$.[10] The introduction of inorganic Cs$^+$ and/or Rb$^+$ cations into FA-rich perovskite compositions has proven to be an efficient strategy to improve the efficiency and stability of perovskite solar cells. Indeed, the introduction of Cs$^+$ into the bulk of the perovskite lattice decreases the amount of phase impurities in the overall perovskite structure and confers higher stability and reproducibility, whereas Rb$^+$ helps to lower non-radiative, trap-assisted recombination.[11,12] Unfortunately, these ions detrimentally change the optoelectronic properties of the perovskite, by increasing its band gap, moving it further away from the optimum Shockley-Queisser band gap for single-junction solar cells.[13]. Moreover, heteroatoms often create inhomogeneities in the perovskite crystal lattice, which are a major source of degradation and promote non-radiative recombination over the long term.[14,15] Apart from heteroatoms, the use of solution-based fabrication methods and the ionic nature of hybrid perovskites causes several drawbacks leading to perovskite films with low phase purity and a high defect density.[16] Defects are preferentially located at the perovskite surface and grain boundaries.[17,18] Vacancies, in particular iodide or cation vacancies, provide diffusion pathways for cations and halide ions, thus, compromising photovoltaic performance and long-term operational stability.[19–29] To mitigate these issues, different strategies have been explored, such as using Lewis acids/bases,[30,31] polymers,[32,33] different organic solvents,[34,35] organic halide salts,[36,37] and ionic liquids.[38,39] However, most passivation research has so far relied on organic salts as passivation agents, while the more stable inorganic ones have been overlooked.

Herein, we investigate the effects of the two most commonly used kinds of inorganic salts, namely caesium halides, CsX, as well as rubidium halides, RbX, (X=I$^-$,Br$^-$,Cl$^-$) as surface passivators instead of bulk additives to mitigate defects at perovskite/HTM interface and enhance device stability. Best devices treated with CsI reached a power conversion efficiency (PCE) of 24.1% with a high fill factor (FF) of 82.2% and 1.16 V open-circuit voltage ($V_{OC}$), whereas control devices delivered a *Voc* of 1.12 V, 80.5% *FF* and 22.73% PCE. CsI-Passivated devices exhibited improved operational stability as compared to the control. Subsequently, we analyzed in-depth the effect of CsI passivation on the structural and optical properties of FAPbI$_3$-rich perovskite by X-ray diffraction (XRD), X-ray photoelectron spectroscopy (XPS), angle-resolved X-ray photoelectron spectroscopy (ARXPS), time-resolved photoluminescence (TRPL), photoluminescence quantum yield (PLQY) determination, time-resolved electroabsorption spectroscopy (TREAS) and Kelvin probe force microscopy (KPFM). Specifically, TREAS, TRPL, and PLQY show that CsI passivation suppresses non-radiative recombination and increases charge carrier mobility.

## Results and Discussion

To begin with, we prepared perovskite precursor solutions based on α-FAPbI$_3$ to which we added 2% of MAPbBr$_3$ to help stabilize the phase by generating a double-cation system, (FAPbI$_3$)$_{0.98}$(MAPbBr$_3$)$_{0.02}$ (control). Based on this, we prepared the target triple- (Cs$_{0.05}$FA$_{0.93}$MA$_{0.02}$Pb(I$_{0.98}$Br$_{0.02}$)$_3$, CsFAMA) and quadruple-cation (Rb$_{0.03}$Cs$_{0.05}$FA$_{0.93}$MA$_{0.02}$Pb(I$_{0.98}$Br$_{0.02}$)$_3$, Rb.CsFAMA) materials by adding the corresponding alkali halide salts into the precursor solution. Perovskite films were deposited by spin-coating onto a mesoporous TiO$_2$ (mp-TiO$_2$) layer using a one-step method employing chlorobenzene as the antisolvent. Initially, we carried out an investigation into the effects of mixed cation- halide on the photovoltaic performance. The champion device using CsFAMA showed a PCE of 22.05% with a $J_{SC}$ of 24.6 mA·cm$^{-2}$, a $V_{OC}$ of 1.13 V and FF of 78.7%, while using Rb.CsFAMA produced a PCE of 22.4% with a $J_{SC}$ of 24.82 mA·cm$^{-2}$, a $V_{OC}$ of 1.16 V and FF of 77.4%. By comparison, a PCE of 22.7% with a $J_{SC}$ of 25.0 mA·cm$^{-2}$, a $V_{OC}$ of 1.12 V and FF of 80.5% was achieved for the best control device as shown in **Figure 1a**.

Subsequently, we excluded the alkali halide salts from the precursor solution, instead employing them as surface passivating agents for the control ((FAPbI$_3$)$_{0.98}$(MAPbBr$_3$)$_{0.02}$) composition. For simplicity, we refer to them as CsX-Passivated and RbX-Passivated films, respectively (X = I, Br, Cl). This strategy led to a dramatic improvement of device performance with a champion PCE of 24.1% ($J_{SC}$ = 25.15 mA·cm$^{-2}$, $V_{OC}$ = 1.164 V, FF = 82.2%) obtained following passivation with CsI whereas RbI passivation resulted in a PCE of 22.95% ($J_{SC}$ = 24.9 mA·cm$^{-2}$, $V_{oc}$ = 1.15 V, FF = 80.0%). These results are displayed in **Figure 1b**, **Table S1 and Figure S1-S4**. To be noted, the control composition we used was firstly treated with ethylammonium iodide (EAI) in order to protect the perovskite film against the harsh methanol solvent in which CsX and RbX salts were dissolved. The role and effect of such method was investigated in **Figure S5** and **Table S2**. It is clear that the double treatment (1$^{st}$ EAI, 2$^{nd}$ CsI) leads to the best efficiency, as the perovskite of EAI-passivated films (control) is protected from the harsh MeOH solvent.

To gain insight into the band gap of perovskite films prepared following these two different strategies, we recorded the incident photon-to-current efficiency (IPCE) spectra, integrated current density and the inflection point (**Figure 1c,d** and **Figure S6a,b**) for control, CsFAMA, Rb.CsFAMA, CsI-Passivated and RbI-Passivated devices. Control, CsI-Passivated and RbI-Passivated films exhibited an identical bandgap of 1.52 eV (**Figure 1d**), whereas CsFAMA and Rb.CsFAMA showed a bandgap of 1.55 eV. Interestingly, and as shown by the onset of the IPCE spectra (**Figure 1c**), $Cs^+$ ions applied via surface passivation did not change the band gap compared to an untreated control device. This suggests that $Cs^+$ does not diffuse into the bulk of the perovskite film but rather acts as a surface defect passivator which is in good agreement with the angle resolved XPS (ARXPS) spectra shown in **Figure S7**.

In an ARXPS experiment, the sample is sequentially tilted with respect to the analyzer. In turn - due to a geometrical effect - the effective escape depth of the photoelectrons decreases as the tilt angle increases. This implies that the contribution of species over-represented at the surface proportionally increases with the tilt angle, with respect to species dominant in sub-surface layers (or uniformly distributed throughout the film). A comparison of the relative abundance of elements at different angles confirmed that $Cs^+$ was concentrated at the surface and not incorporated into the bulk of the perovskite.

After studying the effects of CsI and RbI as bulk-additives as well as surface passivators, we can clearly see that the best results were obtained for CsI-Passivated devices. Therefore, going forward we performed an in-depth investigation on the impact of passivation with CsI using multiple characterization techniques to unravel its effect on devices' photophysics and determine the impact on device stability. In addition, the statistical distribution of photovoltaic characteristics ($J_{SC}$, $V_{OC}$, FF, and PCE) is presented in **Figure 1e** for control and CsI-Passivated devices. Statistical analysis of photovoltaic parameters of CsI-Passivated devices, highlight a clear improvement and reproducible *Voc* and *FF*. Hysteresis behaviour was examined by performing backward and forward *J-V* scans (**Figure S8a,b,c**), revealing hysteresis indices (HI = [(PCE$_{backward}$ − PCE$_{forward}$)/(PCE$_{backward}$)]x100) of 3.17%, 4.70% and 2.28% for a control, RbI-Passivated and CsI-Passivated device, respectively. Shelf-life tests were conducted on Control and CsI-Passivated devices by storing them at room temperature in the dark at 20% RH. The Control and CsI-Passivated devices retained 77.5% and 93% of their initial performance, respectively, after 400 h (Figure S9). Furthermore, we carried out an operational stability test on a control and a CsI-Passivated device by subjecting them to MPP tracking for 600 h under continuous 1 sun illumination at room temperature in a nitrogen environment. The control and CsI-Passivated device retained 80 % and 90 % of their respective initial performance as shown in **Figure 1f**.

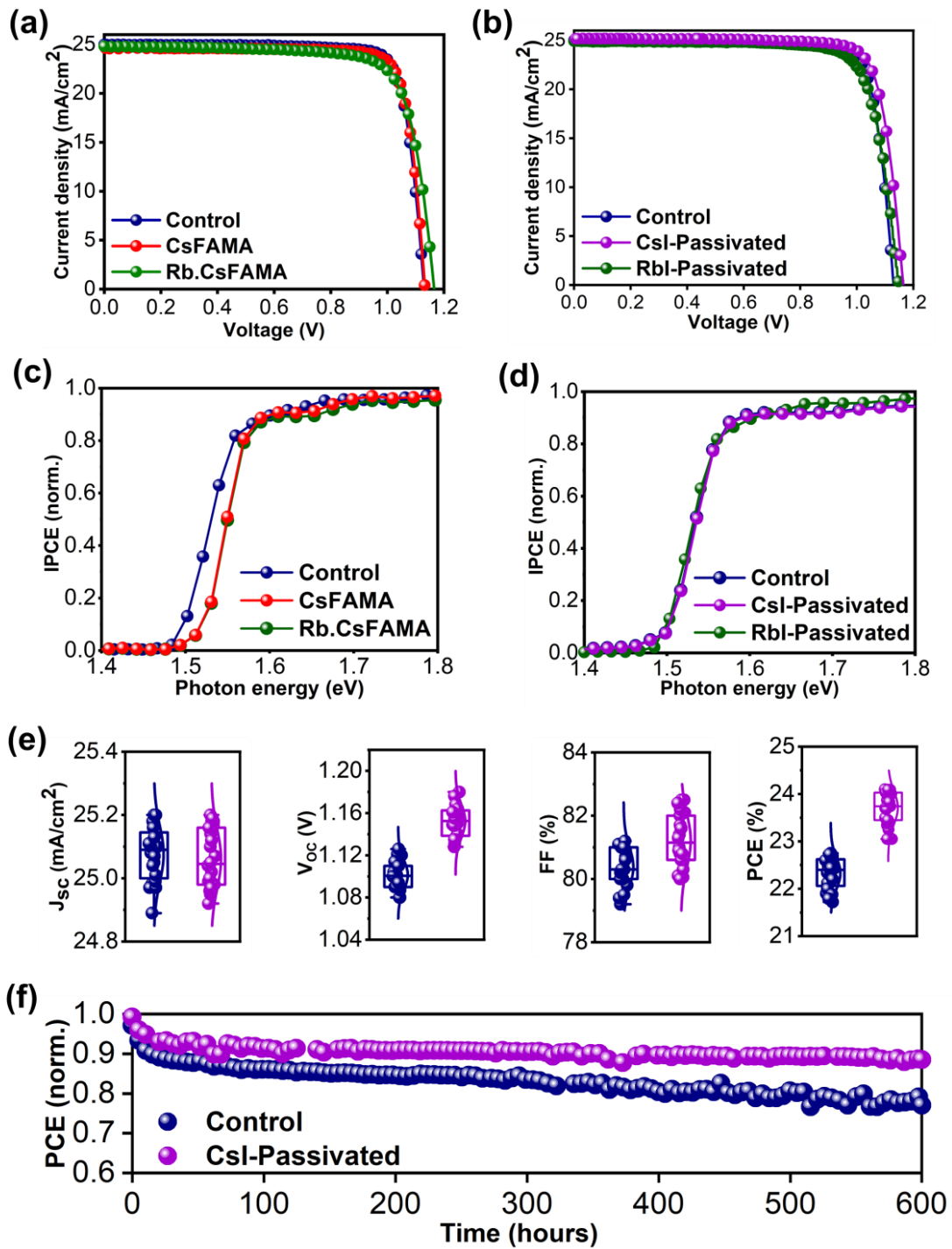

**Figure 1.** (a+b) *J-V* curves of control, CsFAMA, and Rb.CsFAMA (a), as well as CsI-Passivated, and RbI-Passivated (b) devices. (c+d) Inflection points extracted from IPCE spectra of control, CsFAMA, and Rb.CsFAMA (c), as well as CsI-Passivated, and RbI-Passivated (d) devices. (e) Statistical analysis of photovoltaic parameters control and CsI-Passivated devices. (f) Evolution of the normalized PCE of a control and a CsI-Passivated device during MPP tracking in a nitrogen environment at room temperature under continuous irradiation with simulated AM1.5 sunlight.

To analyze the influence of CsI passivation on the structure of the perovskite absorber layer, XRD measurements were conducted (**Figure 2a**). CsI passivation was found to completely remove the signal of $PbI_2$ at 12.6° in the XRD pattern. This suggests full conversion of the reactants into the perovskite phase. Also, in order to probe whether undesired perovskite phases were formed upon the passivation with $CsPbI_3$ (such as α or/ and δ $CsPbI_3$ phases), we performed XRD measurements of the perovskite films at different temperatures (RT, 70 °C, 120 °C and 160 °C as shown in Figure S10) after the CsI post-treatment as shown in **Figure 10.** From the XRD results, we clearly did not observe the formation of such phases. Furthermore, X-ray photoelectron spectroscopy (XPS) measurements (**Figure 2b**) demonstrated that upon the addition of passivating CsI, any residual metallic lead $Pb^0$, which is often observed in MHPs, and is seen in the control films, is completely transformed into $Pb^{2+}$.[40] This is a key finding, as metallic lead aggregates form deep-level traps, a known source of non-radiative recombination and degradation.[41] **Figure 3a** shows the surface morphology of both control and CsI-Passivated perovskite films. CsI-Passivated samples are uniform and highly crystalline with substantially larger perovskite grains as compared to the non-treated perovskite (control). In **Figure 3b** and **Figure S11**, Kelvin probe force microscopy (KPFM) measurements of control and CsI-Passivated films demonstrate the striking effect of CsI in mitigating surface potential inhomogeneities of the perovskite films. CsI passivation yielded films showing a homogeneous surface with potential variations >10 mV, whereas the untreated films (control) exhibited differences close to 30 mV. Such regulation of the surface potential was recently reported to enhance the Quasi-Fermi level splitting (and thus, the $V_{OC}$ of resulting devices) via lowering non-radiative recombination, and indeed, we observed and increase in $V_{OC}$ in the CsI-passivated. Moreover, atomic force microscopy (AFM) measurements revealed that both conditions result in very similar topography as shown in **Figure S12.**

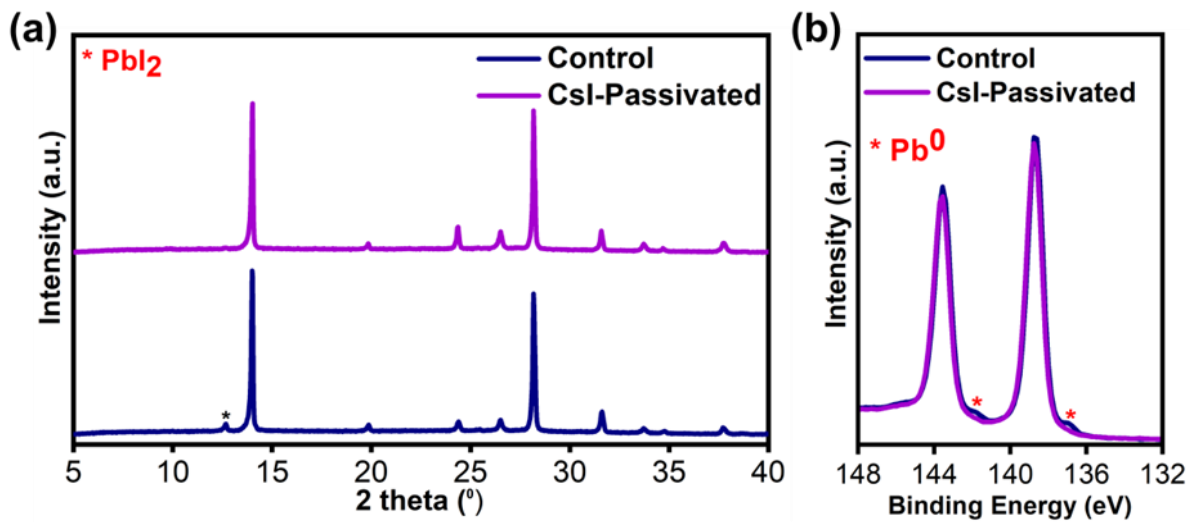

**Figure 2.** (a) XRD patterns of control and CsI-Passivated perovskite films on FTO highlighting the suppression of PbI$_2$ signal upon CsI passivation. (b) XPS Pb 4f spectra of Control and CsI-Passivated showing the suppression of residual unreacted Pb$^0$ upon CsI passivation.

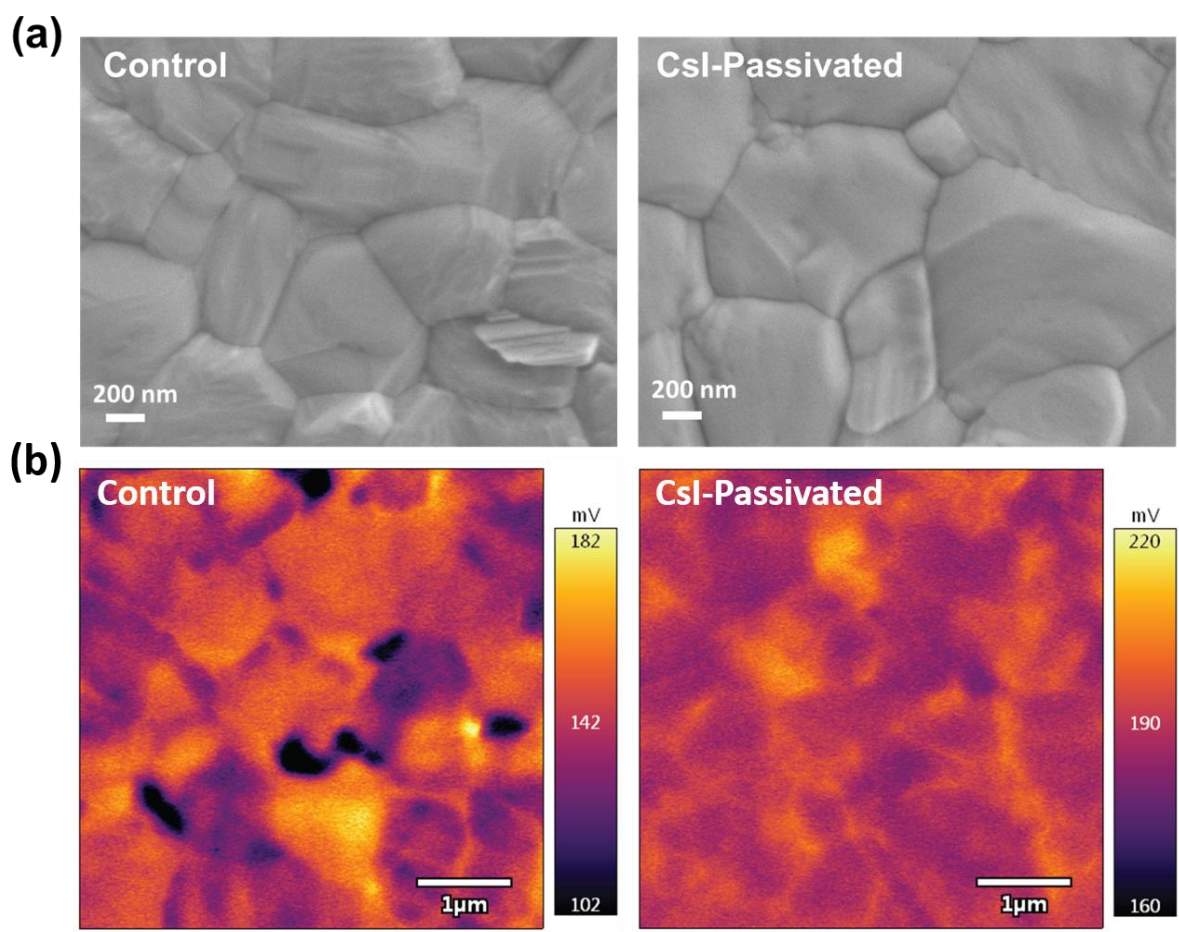

**Figure 3** (a) Top-view SEM images of control and CsI-Passivated perovskite films exhibiting substantially larger grain sizes (around 1 μm) for the latter. (b) KPFM measurements of control and CsI-Passivated films showing a considerably more homogeneous surface potential for the latter.

Next, we carried out an in-depth investigation of the optoelectronic properties of CsI-Passivated perovskite films. We performed time-resolved electroabsorption spectroscopy (TREAS) to get insights into the intrinsic charge carrier mobility and lifetime. Traces of both, control and CsI-Passivated, samples displayed two distinct features at early and late time regimes, respectively (**Figure 4a** and **Figure S13,S14**). The first part consists of a linear increase corresponding to the intragrain charge carrier separation and shows a similar time constant for both conditions.[42] This is followed by a second, logarithmic signal increase caused by the accumulation of charge at the perovskite surface, which can thus be used to determine charge carrier mobilities in these perovskite films. For this part, time constants of $\tau = 2.57$ ps and $\tau = 2.10$ ps were recorded for the control and CsI-Passivated films, respectively. With charge mobility being inversely proportional to this time constant ($\mu = \frac{l}{\tau [E]_0}$; where $\mu$ is the mobility, $l = (l_{e\text{-}} + l_{h\text{+}})/2$ the charge transit distance, averaged here to half of the perovskite layer thickness, $\tau$ the logarithmic domain time constant and $[E]_0$ the bias applied accounting for the voltage drop across the $Al_2O_3$ insulating layer),[42] TREAS demonstrates that CsI passivation leads to a higher mobility of charges, with a calculated mobility of 24.2 $cm^2V^{-1}s^{-1}$ for CsI-Passivated vs. 19.8 $cm^2V^{-1}s^{-1}$ for control films. We assumed equal mobility for holes and electrons. The lifetimes were calculated via TRPL as $1.05 \cdot 10^3$ ns and $1.0 \cdot 10^3$ ns for the control and CsI-Passivated films, respectively, which is in good agreement with TREAS observations. To further verify this, we analyzed the diode characteristics of our devices. In **Figure 4c**, we measured the ideality factor ($n_{id}$) by determining $V_{OC}$ as a function of incident light intensity. Upon CsI treatment $n_{id}$ decreased from 1.54 (control) to 1.27 (CsI-Passivated), which indicates a substantial reduction of non-radiative recombination. To further confirm the role of nonradiative recombination, we measured absolute photoluminescence (PL) photon fluxes $\Phi_{PL}(E)$ of Control and CsI-Passivated films following established methods.[43] Stunningly, the PLQY improved from 0.4% for the untreated perovskite films to as high as 8% for CsI-Passivated films as shown in **Figure 4d** accompanied by an increased Quasi-Fermi level splitting (QFLS) of 1.20 V compared to 1.13 V for the control film. In short, measurements of $n_{id}$, TREAS, TRPL and PLQY indicated that CsI passivation improves the quality of perovskite and reduces the non-radiative recombination rate while increasing the charge carrier mobility, leading to the observed improvements in both $V_{OC}$ and FF.

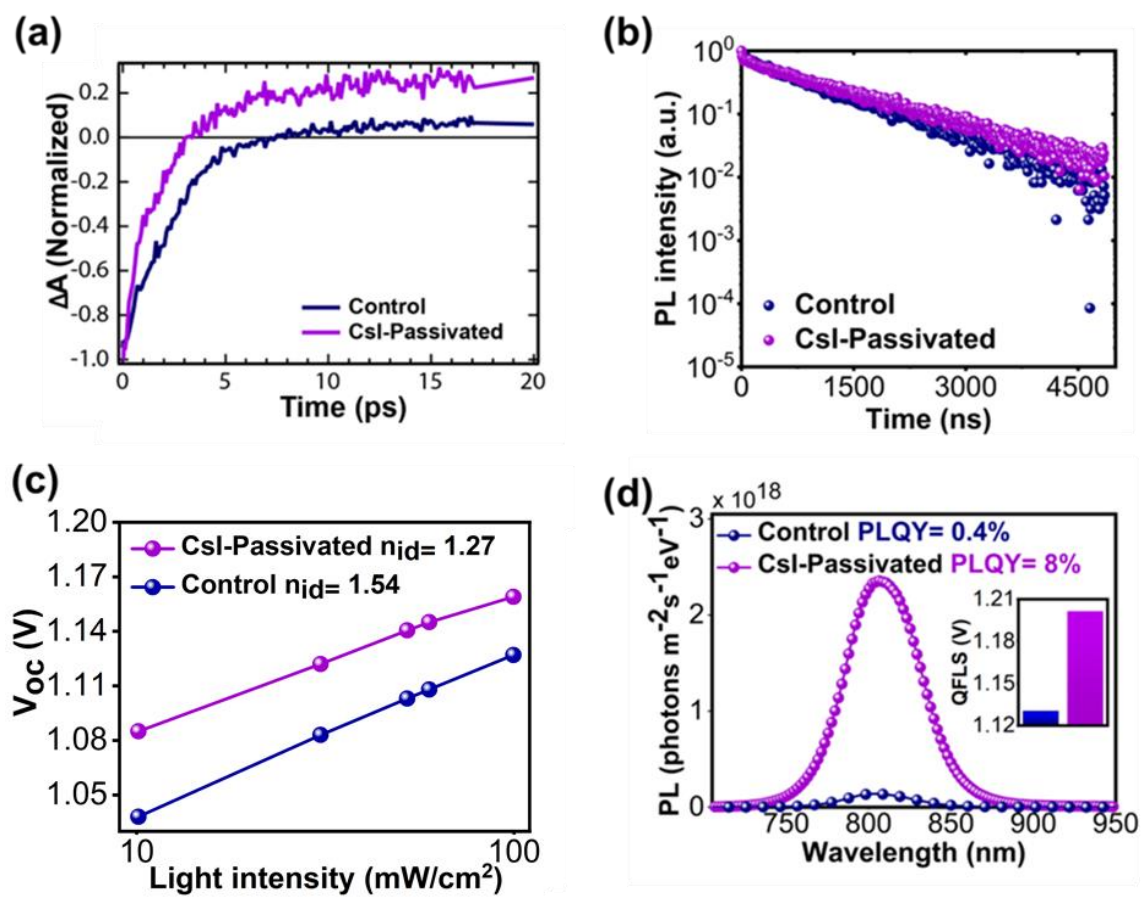

**Figure 4.** (a) TREAS measurements of the control and CsI-Passivated films on TiO$_2$. The higher free charge carrier mobility for CsI-Passivated films (24.2 vs 19.8 cm$^2$V$^{-1}$s$^{-1}$) suggests lower defects. (b) TRPL measurements of the control and CsI-Passivated films on glass. Charge carrier lifetime is improved in CsI-treated films compared to the control. (c) Ideality factor of control and CsI-Passivated PSCs. (d) PLQY measurements of control and CsI-Passivated films exhibiting a PLQY of 0.4 and 8% respectively, showing an impressive 18x increase upon CsI passivation. inset: Quasi-Fermi level splitting of the control and CsI-treated films. The larger QFLS of the CsI-Passivated films attests to their higher quality.

## Conclusion

In summary, we applied CsI as a surface passivator instead of A-cation stoichiometric stabilizer to suppress defects at the interface of perovskite/HTM to obtain highly efficient and stable PSCs based on α-FAPbI$_3$. The objective was to remove Cs$^+$ and Rb$^+$ ions from the perovskite composition to prepare stable and robust perovskite solar cells without sacrificing the optimum *J-V* parameters delivered by the α-FAPbI$_3$ perovskite composition. As a result, CsI-Passivated devices achieved a PCE exceeding 24% and also exhibit a remarkable improvement in structural, morphological, and optoelectronic properties as well as device operational stability based on our in-depth experimental investigation.

**Experimental Section/Methods**

**Solar cells preparation:**

Fluorine-doped tin oxide (FTO)-glass substrates (TCO glass, NSG 10, Nippon sheet glass, Japan) were etched using Zn powder and 4 M HCl, cleaned by ultrasonication in Hellmanex (2%, deionized water), rinsed thoroughly with de-ionized water and ethanol, and then treated in oxygen plasma for 15 min. An approximately 30 nm thick blocking layer (TiO$_2$) was deposited on the cleaned FTO by spray pyrolysis at 450 °C using a commercial titanium diisopropoxide bis(acetylacetonate) solution (75% in 2-propanol, Sigma-Aldrich) diluted in anhydrous ethanol (1:9 volume ratio) as precursor and oxygen as a carrier gas. A mesoporous TiO$_2$ layer was deposited by spin-coating a diluted paste (Dyesol 30NRD) in ethanol (1:6.5 wt. ratio) at 5000 rpm for 15 s and sintering at 450°C for 30 min in dry atmosphere. The perovskite films were deposited using a single-step deposition method from the precursor solution, which was prepared in an argon atmosphere by dissolving CsI, RbI, FAI, MABr, PbI$_2$ and PbBr$_2$ in anhydrous dimethylformamide/dimethyl sulfoxide (4:1 (volume ratio)) to achieve the desired compositions: (FAPbI$_3$)$_{0.98}$(MAPbBr$_3$)$_{0.02}$, Cs$_{0.05}$FA$_{0.93}$MA$_{0.02}$Pb(I$_{0.98}$Br$_{0.02}$)$_3$ and Rb$_{0.03}$Cs$_{0.05}$FA$_{0.93}$MA$_{0.02}$Pb(I$_{0.98}$Br$_{0.02}$)$_3$ using a 3% PbI$_2$ excess and 44 mg MACl. The perovskite films were then passivated by dynamically spin-coating (6000 rpm for 45 s) a 3 mg.ml$^{-1}$ solution of EAI (EAI = ethylammonium iodide = CH$_3$CH$_2$NH$_3^+$ I$^-$) in isopropanol. Afterwards, CsX (X = Br$^-$, Cl$^-$, I$^-$) and RbX (X = Br$^-$, Cl$^-$, I$^-$) were deposited on the surface of EAI-passivated perovskite films by spin-coating solutions (in MeOH) following the same procedure. The HTM was deposited by spin-coating at 5000 rpm for 30 s. The HTM was doped with bis(trifluoromethylsulfonyl)imide lithium salt (17.8 µl of a solution of 520 mg LiTFSI in 1 ml of acetonitrile), and 28.8 µl of 4-tert-butylpyridine. Finally, an approximately 80 nm gold (Au) layer was deposited by thermal evaporation.

**Device characterization:**

The current-voltage (*J-V*) characteristics of the perovskite devices were recorded with a digital source meter (Keithley model 2400, USA). A 450 W xenon lamp (Oriel, USA) was used as the light source for photovoltaic (*J-V*) measurements. The spectral output of the lamp was filtered using a Schott K113 Tempax sunlight filter (Präzisions Glas & Optik GmbH, Germany) to reduce the mismatch between the simulated and actual solar spectrum to less than 2%. The photo-active area of 0.158 cm$^2$ was defined using a dark-colored metal mask.

**Incident photon-to-current efficiency (IPCE):**

IPCE was recorded under a constant white light bias of approximately 5 mW cm$^{-2}$ supplied by an array of white light emitting diodes. The excitation beam coming from a 300 W Xenon lamp (ILC Technology) was focused through a Gemini- 180 double monochromator (Jobin Yvon Ltd) and chopped at approximately 2 Hz. The signal was recorded using a Model SR830 DSP Lock-In Amplifier (Stanford Research Systems).

**Scanning electron microscopy (SEM):**

SEM micrograph measurements were performed on a ZEISS Merlin HR-SEM.

**X-ray powder diffraction (XRD):**

XRD patterns of the perovskite films were recorded on an X'Pert MPD PRO (Panalytical) equipped with a ceramic tube (Cu anode, $\lambda$ = 1.54060 Å), a secondary graphite (002) monochromator and a RTMS X'Celerator (Panalytical).

**Atomic force microscopy (AFM) and kelvin probe force microscopy (KPFM):**

AFM/KPFM studies were performed with an Asylum Research Cypher using Pt coated tips (AC240TM, Olympus) under ambient condition at room temperature. Films were measured on the grounded FTO substrate for KPFM characterizations. The resulting images were processed by flattening.

**Photoluminescence quantum yield (PLQY):**

The PLQY was acquired following the procedure suggested by de Mello.[43] Samples were excited using a continuous-wave laser (OBIS LX, 660 nm) whose power was adjusted to match the photogeneration rate under 1 sun of illumination (0.324 mW, 0.786 mm effective beam full width at half-maximum (fwhm)). The signal was collected using an integrating sphere (Gigahertz Optik, UPB-150-ARTA) connected via a multimode, 400 μm diameter optical fiber (Thorlabs BFL44LS01) to a spectrometer (Andor, Kymera 193i). The system was spectrally calibrated using an irradiance calibration standard lamp (Gigahertz Optik, BN-LH250-V01).

**Time-resolved photoluminescence (TRPL):**

The TRPL of perovskite films on glass was measured with a spectrometer (FluoroLog-3, Horiba) working in a time-correlated single-photon counting mode with <ns time resolution. A picosecond pulsed diode laser head NanoLED N-670L (Horiba) emitting <200 ps duration pulses at 670 nm with a maximum repetition rate of 1 MHz was used as excitation source.

**Electroabsorption (EA) and time-resolved electroabsorption spectroscopy (TREAS) measurements:**

EA and TREAS measurements were carried out using a femtosecond pump probe spectrometer based on a Ti:sapphire laser (Clark-MXR, CPA-2001) delivering 778 nm pulses with a pulse duration of 150 fs and a 1 kHz repetition rate. The difference between the two techniques is that TREAS involves pumping the sample, whereas EA just uses the probe beam. The 389 nm pump beam for TREAS measurements was generated by passing part of the fundamental through a beta barium borate (BBO) crystal in order to generate the second harmonic at 389 nm. In both, EA and TREAS, the probe beam, consisting of a broadband white light continuum (400–750 nm), was generated by passing part of the fundamental through a 4 mm sapphire window. Measurements were carried out in reflection mode, and the probe beam was split into a signal and reference beam to account for shot-to-shot fluctuations. The signal and reference beams were dispersed into two spectrographs (SR163, Andor Instruments) and detected shot-to-shot at 1 kHz using 512x68 pixel back-thinned CCD cameras (Hamamatsu S07030-0906). Square pulses generated by a function generator (Tektronix AFG 2021, from –10 to 10 V, 100 µs pulse duration) were used to modulate the electric field across the sample at 500 Hz. The current responses across the samples were recorded using a 50 Ω series load with a 400 MHz bandpass oscilloscope (Tektronix TDS 3044B).

**X-ray photoelectron spectroscopy (XPS and Angle-Resolved X-ray photoelectron spectroscopy (ARXPS):**

XPS and ARXPS measurements were carried out using a PHI VersaProbe II scanning XPS microprobe with an Al Kα X-ray source. For XPS measurements, the spherical capacitor analyser was set at 45° take-off angle with respect to the sample surface. Bulk composition analysis was done after argon plasma etching. Data were processed using the PHI Multipak software.

**Long term light soaking test:**

Stability measurements were performed with a Biologic MPG2 potentiostat under a full AM 1.5 Sun-equivalent white LED lamp. The devices were measured with a maximum power point (MPP) tracking routine under continuous illumination at room temperature. The MPP was updated every 10 s by a standard perturb and observe method. Every minute a *J-V* curve was recorded in order to track the evolution of individual *J-V* parameters.

**Acknowledgements**

T.P.B. gratefully acknowledges support from the Swiss National Science Foundation (project no. IZJSZ2_180176). E.A.A, M.T.A, M.S.A and A.S.A gratefully acknowledges the support of King Abdulaziz City for Science and Technology (KACST), Saudi Arabia. G.K. acknowledges financial support from the European Commission (Marie Skłodowska-Curie Actions, H2020-MSCA-IF-2020, project id 101024237). G.C.F and J.-E.M thank the Swiss National Science Foundation (Grant No. 200021_175729) and NCCR-MUST for financial support. J.-H.Y gratefully acknowledges the support of the Korea Electric Power Corporation (KEPCO) collaboration project.

**Author Contributions**

E.A.A conceived the idea of the work, designed and planned the experiments. T.B., M.T.A and M.S.A fabricated and optimized the perovskite solar cell devices, did all the basic characterizations, analyzed the data and wrote the manuscript with support from E.A.A. G.K. contributed towards device fabrication. M.T.A, M.S.A and A.S.A contributed to stability and XRD measurements. G.C.F. and J.E.M. were responsible for EA and TREAS analysis. M.D. and J.Y performed the XPS/ARXPS and the AFM/KPFM measurements, respectively. SEM micrographs were recorded by J.G with help of T.B. B.C. and F.E recorded PLQY and TRPL spectra and analyzed the data. LP was involved in discussing the obtained results. M.G. , S.M.Z. and E.A.A supervised the project and participated in discussing the results. All authors contributed towards the preparation of the manuscript.

# References

[1] M. Grätzel, The light and shade of perovskite solar cells, *Nat. Mater.* **2014**, *13*, 838.

[2] N. G. Park, M. Grätzel, T. Miyasaka, K. Zhu, K. Emery, Towards stable and commercially available perovskite solar cells, *Nat. Energy* **2016**, *1*, 1.

[3] J. P. Correa-Baena, M. Saliba, T. Buonassisi, M. Grätzel, A. Abate, W. Tress, A. Hagfeldt, Promises and challenges of perovskite solar cells, *Science.* **2017**, *358*, 739.

[4] M. A. Green, A. Ho-Baillie, H. J. Snaith, The emergence of perovskite solar cells, *Nat. Photonics* **2014**, *8*, 506.

[5] A. Kojima, K. Teshima, Y. Shirai, T. Miyasaka, rganometal Halide Perovskites as Visible-Light Sensitizers for Photovoltaic Cells, *J. Am. Chem. Soc.* **2009**, *131*, 6050.

[6] National Renewable Energy Laboratory of the United States of America (NREL) National Center for Photovoltaics, "Best Research-Cell Effficiencies chart," can be found under https://www.nrel.gov/pv/assets/pdfs/best-research-cell-efficiencies.20200925.pdf, **n.d.**

[7] Q. Han, S. H. Bae, P. Sun, Y. T. Hsieh, Y. Yang, Y. S. Rim, H. Zhao, Q. Chen, W. Shi, G. Li, Y. Yeng, Single Crystal Formamidinium Lead Iodide (FAPbI3): Insight into the Structural, Optical, and Electrical Properties, *Adv. Mater.* **2016**, *28*, 2253.

[8] G. E. Eperon, S. D. Stranks, C. Menelaou, M. B. Johnston, L. M. Herz, H. J. Snaith, Formamidinium lead trihalide: a broadly tunable perovskite for efficient planar heterojunction solar cells, *Energy Environ. Sci.* **2014**, *7*, 982.

[9] A. Amat, E. Mosconi, E. Ronca, C. Quarti, P. Umari, M. K. Nazeeruddin, M. Grätzel, F. De Angelis, Cation-Induced Band-Gap Tuning in Organohalide Perovskites: Interplay of Spin–Orbit Coupling and Octahedra Tilting, *Nano Lett.* **2014**, *14*, 3608.

[10] E. Smecca, Y. Numata, I. Deretzis, G. Pellegrino, S. Boninelli, T. Miyasaka, A. La Magna, A. Alberti, Stability of solution-processed MAPbI3 and FAPbI3 layers, *Phys. Chem. Chem. Phys.* **2016**, *18*, 13413.

[11] M. Saliba, T. Matsui, J. Y. Seo, K. Domanski, J. P. Correa-Baena, M. K. Nazeeruddin, S. M. Zakeeruddin, W. Tress, A. Abate, A. Hagfeldt, M. Grätzel, Cesium-containing triple cation perovskite solar cells, *Energy Environ. Sci.* **2016**, *9*, 1989.

[12] P. Yadav, M. I. Dar, N. Arora, E. A. Alharbi, F. Giordano, S. M. Zakeeruddin, M. Grätzel, The Role of Rubidium in Multiple-Cation-Based High-Efficiency Perovskite Solar Cells, *Adv. Mater.* **2017**, *29*, 1701077.

[13] W. Shockley, H. J. Queisser, R. ell, Detailed Balance Limit of Efficiency of p-n Junction Solar Cells, *Cit. J. Appl. Phys* **1961**, *32*, 510.

[14] J. P. Correa-Baena, Y. Luo, T. M. Brenner, J. Snaider, S. Sun, X. Li, M. A. Jensen, N. T. P. Hartono, L. Nienhaus, S. Wieghold, J. R. Poindexter, S. Wang, Y. S. Meng, T. Wang, B. Lai, M. V. Holt, Z. Cai, M. G. Bawendi, L. Huang, T. Buonassisi, D. P. Fenning, Homogenized halides and alkali cation segregation in alloyed organic-inorganic perovskites., *Science.* **2019**, *363*, 627.

[15] D. J. Kubicki, D. Prochowicz, A. Hofstetter, S. M. Zakeeruddin, M. Grätzel, L. Emsley, Phase Segregation in Cs-, Rb- and K-Doped Mixed-Cation (MA)x(FA)1–xPbI3 Hybrid Perovskites from Solid-State NMR, *J. Am. Chem. Soc.* **2017**, *139*, 14173.

[16] M. Jung, S. G. Ji, G. Kim, S. Il Seok, Perovskite precursor solution chemistry: from fundamentals to photovoltaic applications, *Chem. Soc. Rev.* **2019**, *48*, 2011.

[17] Q. Jiang, Y. Zhao, X. Zhang, X. Yang, Y. Chen, Z. Chu, Q. Ye, X. Li, Z. Yin, J. You, Surface passivation of perovskite film for efficient solar cells, *Nat. Photonics* **2019**, *13*, 460.

[18] Y. Yang, S. Feng, W. Xu, M. Li, L. Li, X. Zhang, G. Ji, X. Zhang, Z. Wang, Y. Xiong, L. Cao, B. Sun, X. Gao, Enhanced Crystalline Phase Purity of CH3NH3PbI3-xClx Film for High-Efficiency Hysteresis-Free Perovskite Solar Cells *ACS Appl. Mater. Interfaces* **2017**, *9*, 23141.

[19] N. G. Park, K. Zhu, Scalable fabrication and coating methods for perovskite solar cells and solar modules, *Nat. Rev. Mater.* **2020**, *5*, 333.

[20] M. Saliba, T. Matsui, K. Domanski, J. Y. Seo, A. Ummadisingu, S. M. Zakeeruddin, J. P. Correa-Baena, W. R. Tress, A. Abate, A. Hagfeldt, M. Grätzel, Incorporation of rubidium cations into perovskite solar cells improves photovoltaic performance., *Science.* **2016**, *354*, 206.

[21] R. J. Stoddard, A. Rajagopal, R. L. Palmer, I. L. Braly, A. K. Y. Jen, H. W. Hillhouse, Enhancing Defect Tolerance and Phase Stability of High-Bandgap Perovskites via Guanidinium Alloying. *ACS Energy Lett.* **2018**, *3*, 1261.

[22] S. H. Turren-Cruz, A. Hagfeldt, M. Saliba, Methylammonium-free, high-performance, and stable perovskite solar cells on a planar architecture., *Science.* **2018**, *362*, 449.

[23] M. M. Tavakoli, P. Yadav, D. Prochowicz, M. Sponseller, A. Osherov, V. Bulović, J. Kong, ontrollable Perovskite Crystallization via Antisolvent Technique Using Chloride Additives for Highly Efficient Planar Perovskite Solar Cells, *Adv. Energy Mater.* **2019**, *9*, DOI 10.1002/aenm.201803587.

[24] L. Shi, M. P. Bucknall, T. L. Young, M. Zhang, L. Hu, J. Bing, D. S. Lee, J. Kim, T. Wu, N.

Takamure, D. R. McKenzie, S. Huang, M. A. Green, A. W. Y. Ho-Baillie, Gas chromatography-mass spectrometry analyses of encapsulated stable perovskite solar cells, *Science.* **2020**, *368*, DOI 10.1126/science.aba2412.

[25] Y. Yun, F. Wang, H. Huang, Y. Fang, S. Liu, W. Huang, Z. Cheng, Y. Liu, Y. Cao, M. Gao, L. Zhu, L. Wang, T. Qin, W. Huang, A Nontoxic Bifunctional (Anti)Solvent as Digestive-Ripening Agent for High-Performance Perovskite Solar Cells., *Adv. Mater.* **2020**, *32*, DOI 10.1002/adma.201907123.

[26] J. Burschka, N. Pellet, S. J. Moon, R. Humphry-Baker, P. Gao, M. K. Nazeeruddin, M. Grätzel, Sequential deposition as a route to high-performance perovskite-sensitized solar cells., *Nature* **2013**, *499*, 316.

[27] J. Jeong, M. Kim, J. Seo, H. Lu, P. Ahlawat, A. Mishra, Y. Yang, M. A. Hope, F. T. Eickemeyer, M. Kim, Y. Jin Yoon, I. Woo Choi, B. Primera Darwich, S. Ju Choi, Y. Jo, J. Hee Lee, B. Walker, S. M. Zakeeruddin, L. Emsley, U. Rothlisberger, A. Hagfeldt, D. Suk Kim, M. Grätzel, J. Young Kim, Pseudo-halide anion engineering for α-FAPbI3 perovskite solar cells., *Nature* **2021**, *592*, 381.

[28] L. Merten, A. Hinderhofer, T. Baumeler, N. Arora, J. Hagenlocher, S. M. Zakeeruddin, M. I. Dar, M. Grätzel, F. Schreiber, Quantifying Stabilized Phase Purity in Formamidinium-Based Multiple-Cation Hybrid Perovskites, *Chem. Mater.* **2021**, acs. chemmater.0c04185.

[29] X. Zheng, B. Chen, J. Dai, Y. Fang, Y. Bai, Y. Lin, H. Wei, X. C. Zeng, J. Huang, Defect passivation in hybrid perovskite solar cells using quaternary ammonium halide anions and cations *Nat. Energy 2017 27* **2017**, *2*, 1.

[30] Y. Lin, L. Shen, J. Dai, Y. Deng, Y. Wu, Y. Bai, X. Zheng, J. Wang, Y. Fang, H. Wei, W. Ma, X. Cheng Zeng, X. Zhan, J. Huang, Y. Lin, L. Shen, Y. Deng, Y. Bai, X. Zheng, Y. Fang, H. Wei, J. Huang, J. Dai, X. C. Zeng, Y. Wu, W. Ma, J. Wang, X. Zhan, π-Conjugated Lewis Base: Efficient Trap-Passivation and Charge-Extraction for Hybrid Perovskite Solar Cells, *Adv. Mater.* **2017**, *29*, 1604545.

[31] W. Tress, K. Domanski, B. Carlsen, A. Agarwalla, E. A. Alharbi, M. Graetzel, A. Hagfeldt, Performance of perovskite solar cells under simulated temperature-illumination real-world operating conditions, *Nat. Energy* **2019**, *4*, 568.

[32] M. Kim, S. G. Motti, R. Sorrentino, A. Petrozza, Enhanced solar cell stability by hygroscopic polymer passivation of metal halide perovskite thin film., *Energy Environ. Sci.* **2018**, *11*, 2609.

[33] Y. Liu, S. Akin, L. Pan, R. Uchida, N. Arora, J. V. Milić, A. Hinderhofer, F. Schreiber, A. R. Uhl, S. M. Zakeeruddin, A. Hagfeldt, M. I. Dar, M. Grätzel, Ultrahydrophobic 3D/2D

fluoroarene bilayer-based water-resistant perovskite solar cells with efficiencies exceeding 22 *Sci. Adv.* **2019**, *5*, eaaw2543.

[34] R. Wang, J. Xue, L. Meng, J.-W. Lee, Z. Zhao, P. Sun, L. Cai, T. Huang, Z. Wang, Z.-K. Wang, Y. Duan, J. L. Yang, S. Tan, Y. Yuan, Y. Huang, Y. Yang, Caffeine Improves the Performance and Thermal Stability of Perovskite Solar Cells *Joule* **2019**, *3*, 1464.

[35] S. Zhang, S. Wu, W. Chen, H. Zhu, Z. Xiong, Z. Yang, C. Chen, R. Chen, L. Han, W. Chen, Solvent engineering for efficient inverted perovskite solar cells based on inorganic CsPbI2Br light absorber, *Mater. Today Energy* **2018**, *8*, 125.

[36] E. A. Alharbi, A. Y. Alyamani, D. J. Kubicki, A. R. Uhl, B. J. Walder, A. Q. Alanazi, J. Luo, A. Burgos-Caminal, A. Albadri, H. Albrithen, M. H. Alotaibi, J. E. Moser, S. M. Zakeeruddin, F. Giordano, L. Emsley, M. Grätzel, Atomic-level passivation mechanism of ammonium salts enabling highly efficient perovskite solar cells., *Nat. Commun.* **2019**, *10*, 1.

[37] M. Yang, T. Zhang, P. Schulz, Z. Li, G. Li, D. H. Kim, N. Guo, J. J. Berry, K. Zhu, Y. Zhao, Facile fabrication of large-grain CH3NH3PbI3−xBrx films for high-efficiency solar cells via CH3NH3Br-selective Ostwald ripening, *Nat. Commun.* **2016**, *7*, 12305.

[38] S. Bai, P. Da, C. Li, Z. Wang, Z. Yuan, F. Fu, M. Kawecki, X. Liu, N. Sakai, J. T. W. Wang, S. Huettner, S. Buecheler, M. Fahlman, F. Gao, H. J. Snaith, Planar perovskite solar cells with long-term stability using ionic liquid additives., *Nat. 2019 5717764* **2019**, *571*, 245.

[39] J.-Y. Seo, T. Matsui, J. Luo, J.-P. Correa-Baena, F. Giordano, M. Saliba, K. Schenk, A. Ummadisingu, K. Domanski, M. Hadadian, A. Hagfeldt, S. M. Zakeeruddin, U. Steiner, M. Grätzel, A. Abate, J. Seo, J. Luo, F. Giordano, M. Saliba, A. Ummadisingu, K. Domanski, S. M. Zakeeruddin, M. Grätzel, A. Abate, T. Matsui, J. Correa-Baena, M. Hadadian, A. Hagfeldt, K. Schenk, Ionic liquid control crystal growth to enhance planar perovskite solar cells efficiency, *Adv. Energy Mater.* **2016**, *6*, 1600767.

[40] E. A. Alharbi, A. Krishna, T. P. Baumeler, M. Dankl, G. C. Fish, F. Eickemeyer, O. Ouellette, P. Ahlawat, V. Škorjanc, E. John, B. Yang, L. Pfeifer, C. E. Avalos, L. Pan, M. Mensi, P. A. Schouwink, J. E. Moser, A. Hagfeldt, U. Rothlisberger, S. M. Zakeeruddin, M. Grätzel, Methylammonium Triiodide for Defect Engineering of High-Efficiency Perovskite Solar Cells, *ACS Energy Lett.* **2021**, *6*, 3650.

[41] D. Luo, R. Su, W. Zhang, Q. Gong, R. Zhu, Minimizing non-radiative recombination losses in perovskite solar cells, *Nat. Rev. Mater. 2019 51* **2019**, *5*, 44.

[42] A. A. Paraecattil, J. De Jonghe-Risse, V. Pranculis, J. Teuscher, J. E. Moser, Dynamics of Photocarrier Separation in MAPbI3 Perovskite Multigrain Films under a Quasistatic Electric

Field, *J. Phys. Chem. C* **2016**, *120*, 19595.

[43] J. C. De Mello, H. F. Wittmann, R. H. Friend, An improved experimental determination of external photoluminescence quantum efficiency, *Adv. Mater.* **1997**, *9*, 230.

Supporting Information

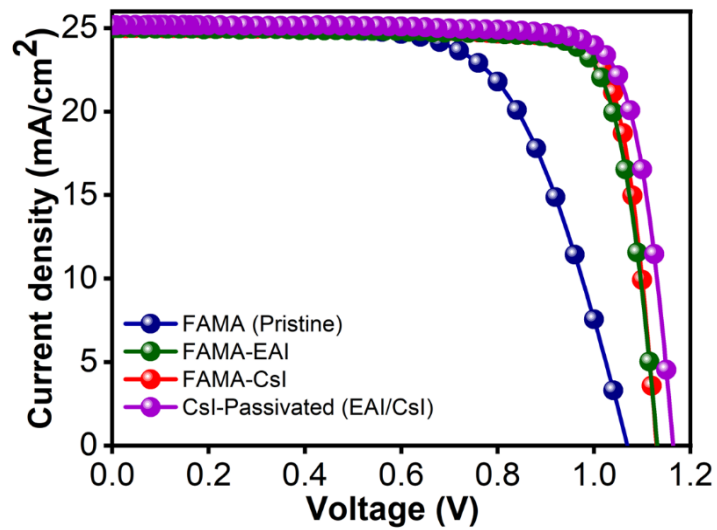

**Figure S1** *J–V* curves of the FAMA, FAMA-EAI, FAMA-CsI and CsI-Passivated (EAI/CsI)-treated perovskite solar cells

**Table S1** PV parameters extracted from Figure S1

|  | $V_{oc}$ (V) | $J_{sc}$ (mA/cm$^2$) | FF(%) | PCE (%) |
|---|---|---|---|---|
| FAMA-Pristine | 1.067 | 25.16 | 75.0 | 20.13 |
| FAMA-EAI | 1.127 | 25.0 | 80.5 | 22.73 |
| FAMA-CsI | 1.130 | 24.96 | 81.6 | 23.06 |
| FAMA – CsI-Passivated (EAI/CsI) | 1.164 | 25.15 | 82.1 | 24.10 |

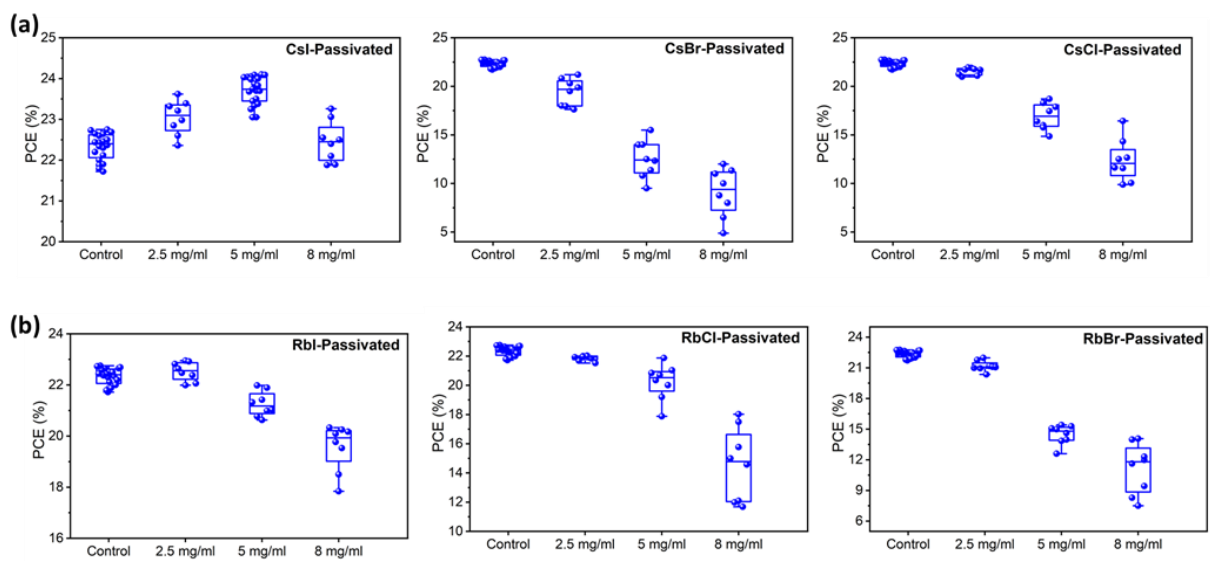

**Figure S2** Device statistics for different CsX, CsRb-Passivated perovskite solar cells

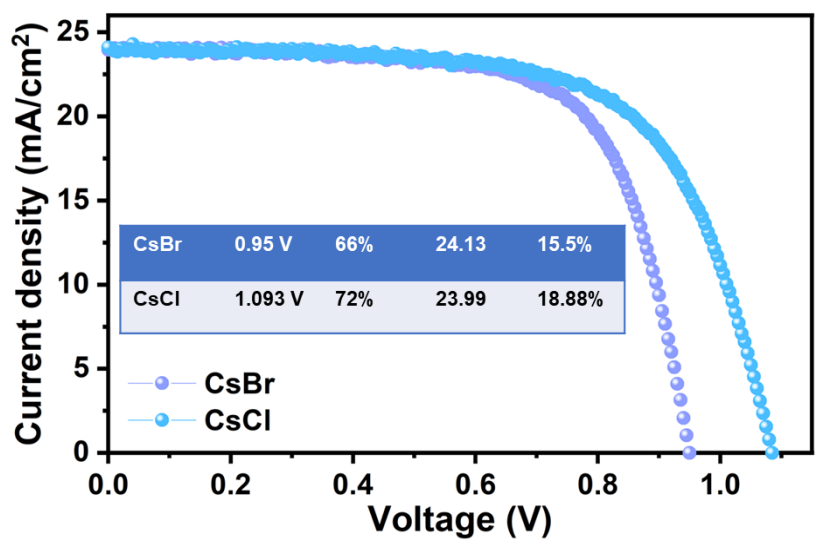

**Figure S3** *J-V* characteristics curves of the control perovskite films passivated with CsX salts (X = Br⁻, Cl⁻)

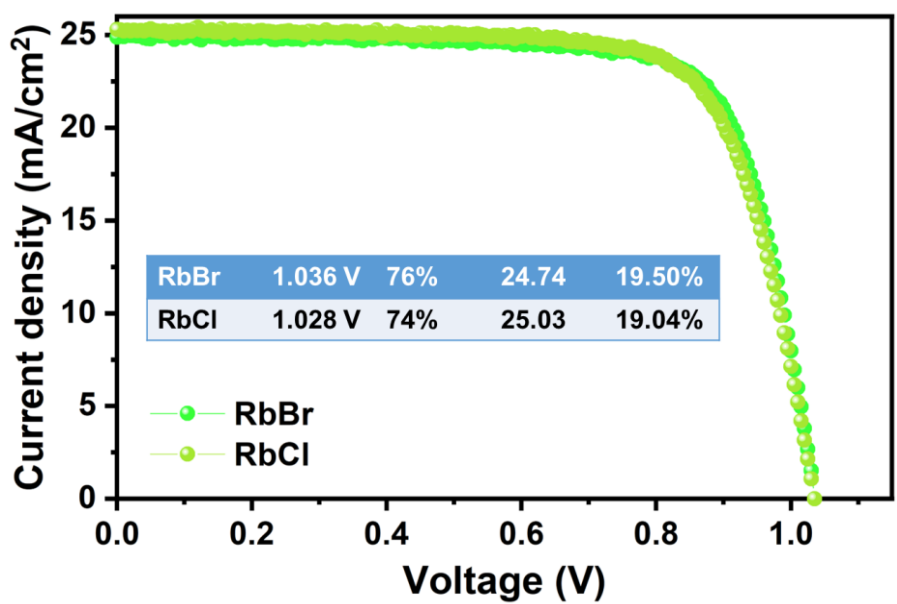

**Figure S4** *J-V* characteristics curves of the control perovskite films passivated with RbX salts (X = Br⁻, Cl⁻) with CsI displayed for comparison

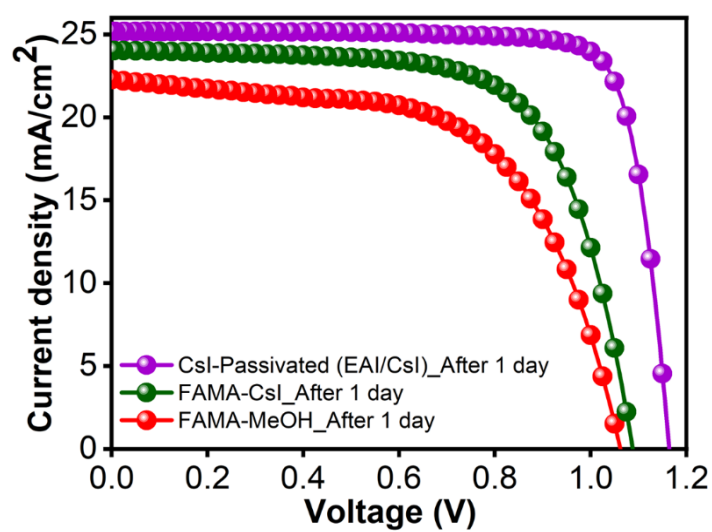

**Figure S5** *J–V* curves of the CsI-Passivated, FAMA-CsI and FAMA-MeOH-treated perovskite solar cells after 1 day outside (ambient stability).

**Table S2** PV parameters extracted from Figure S5

|  | $V_{oc}$ (V) | $J_{sc}$ (mA/cm$^2$) | FF(%) | PCE (%) |
|---|---|---|---|---|
| FAMA-CsI | 1.083 | 24.03 | 68.0 | 17.70% |
| FAMA-MeOH | 1.062 | 22.29 | 60.3 | 14.27% |
| FAMA – CsI-Passivated (EAI/CsI) | 1.164 | 25.15 | 82.1 | 24.10 |

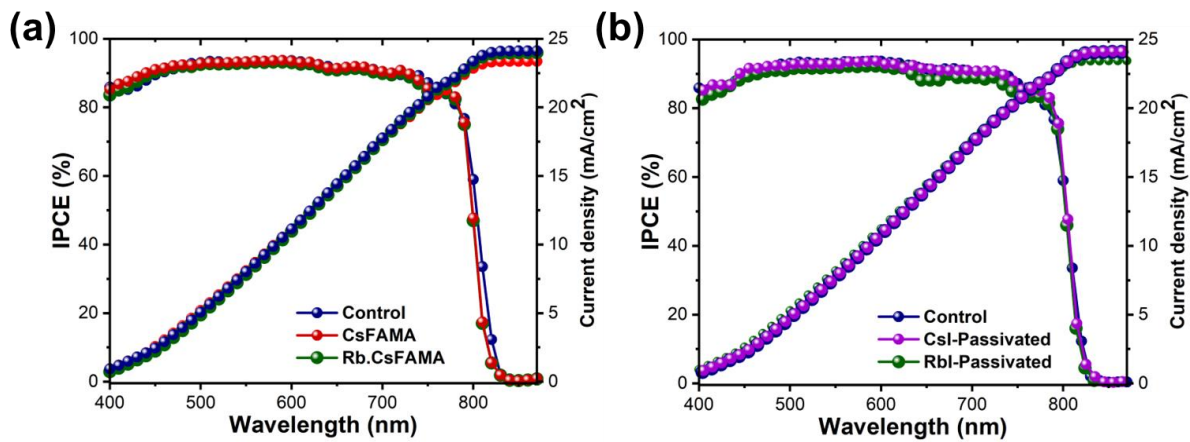

**Figure S6** (a) IPCE spectra of control, CsFAMA and Rb.CsFAMA (a) showing the detrimental effect of integrating Rb and Cs on photovoltaic performance and the typical blue-shift of the absorption of Cs-containing films. (b) IPCE spectra of control, CsI- and RbI-Passivated films showing no shift upon CsI passivation, suggesting no incorporation of $Cs^+$ ions into the perovskite lattice

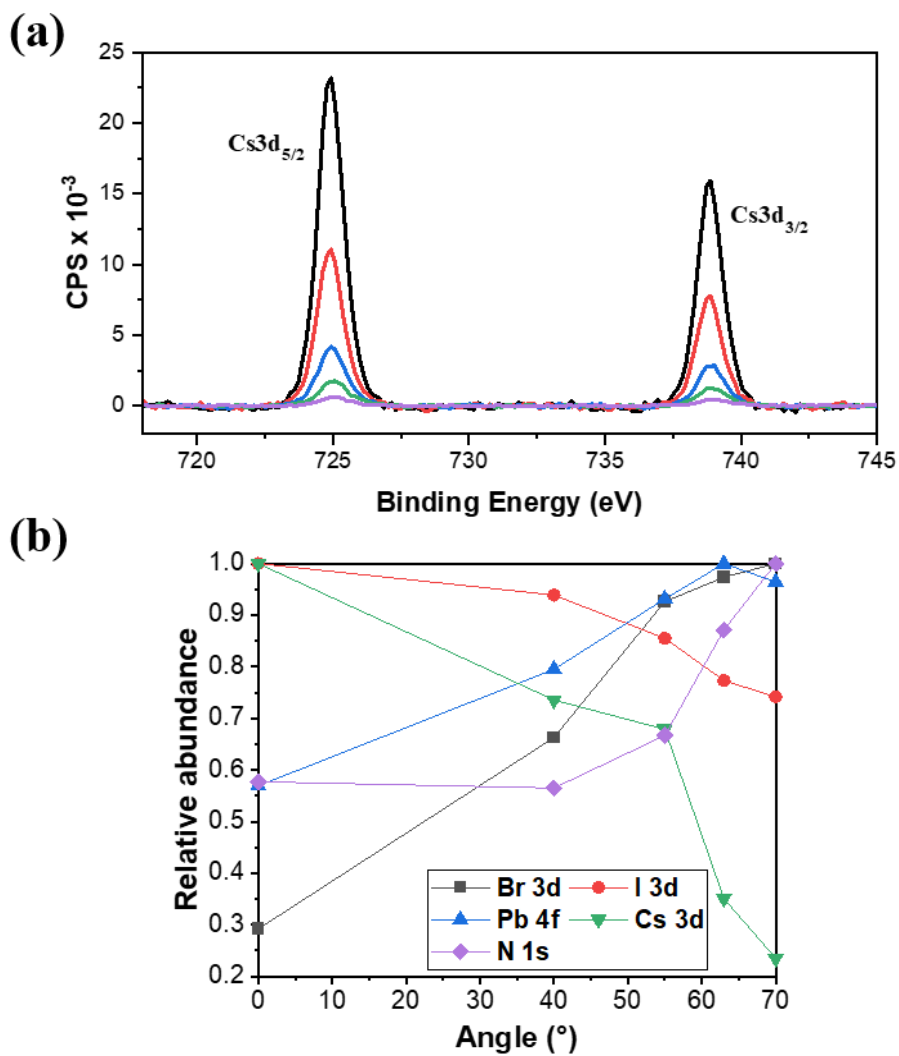

**Figure S7** (a) ARXPS spectra of Cs 3d measured at 0, 40, 55, 63 and 70 degrees (b) relative abundance of Cs(3d), Pb(4f), N(1s), I(3d) and Br(3d) in function of the ARXPS angle showing the gradient of cesium concentration within the first few nanometers of CsI perovskite film. Cs 3d signal intensity decreases as the ARXPS angle is reduced.

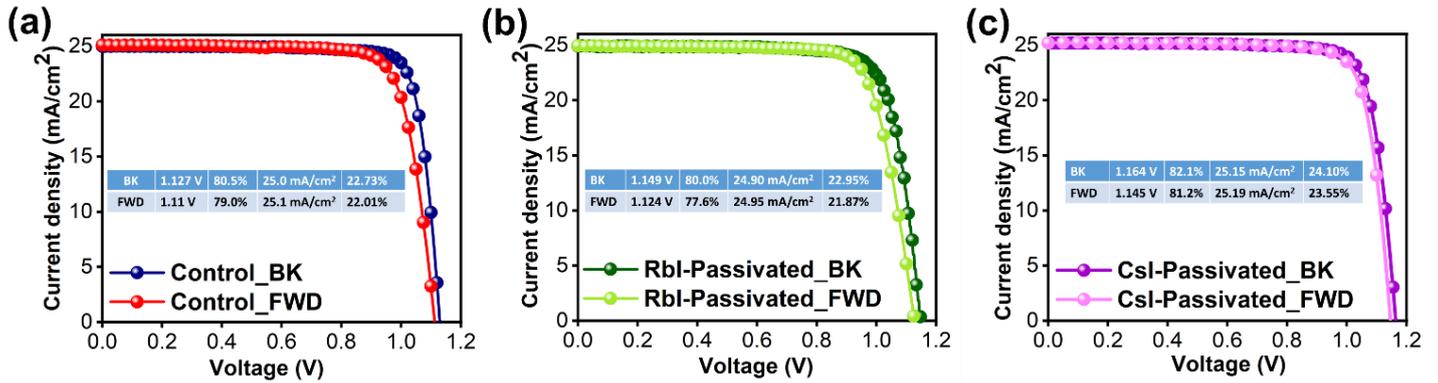

**Figure S8** Backward and forward *J-V* scans of Control (a), RbI-Passivated (b) and CsI-Passivated (c) samples.

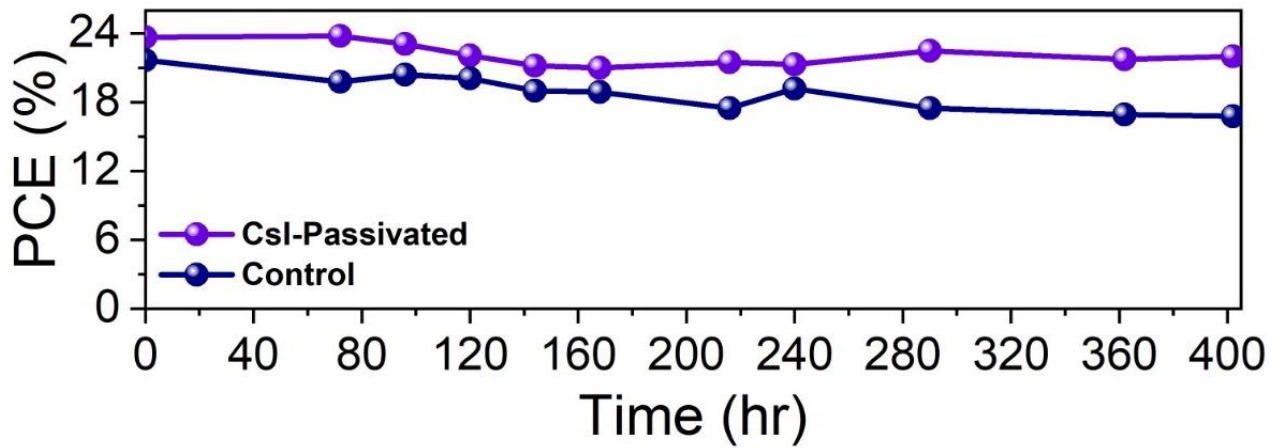

Figure S9. Shelf stability at RT and ~10% RH of control and CsI-Passivated.

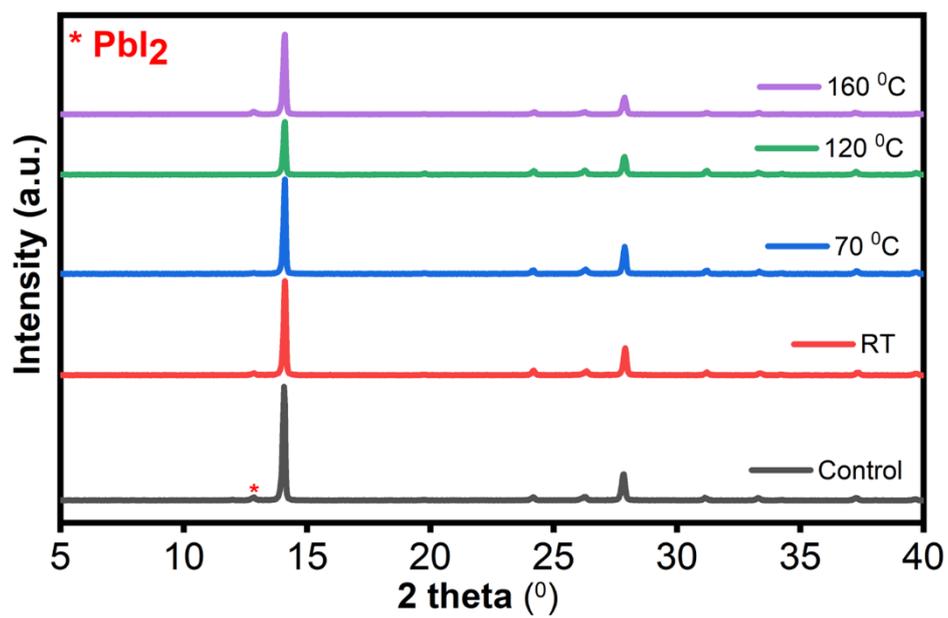

**Figure S10** XRD of perovskite films after CsI post-treatment at different temperatures (RT, 70 °C, 120 °C and 160 °C).

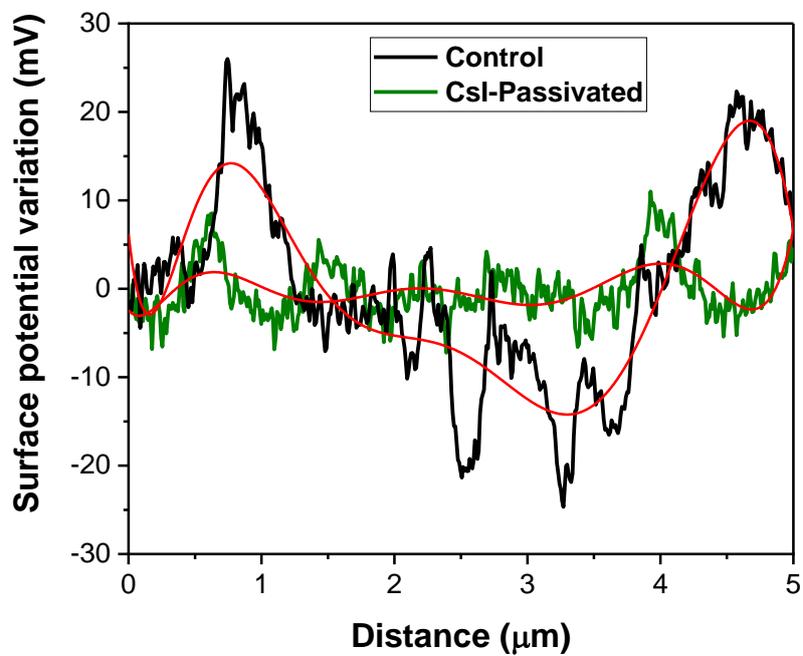

**Figure S11** Surface potential variation of control and CsI-treated films measured by KPFM

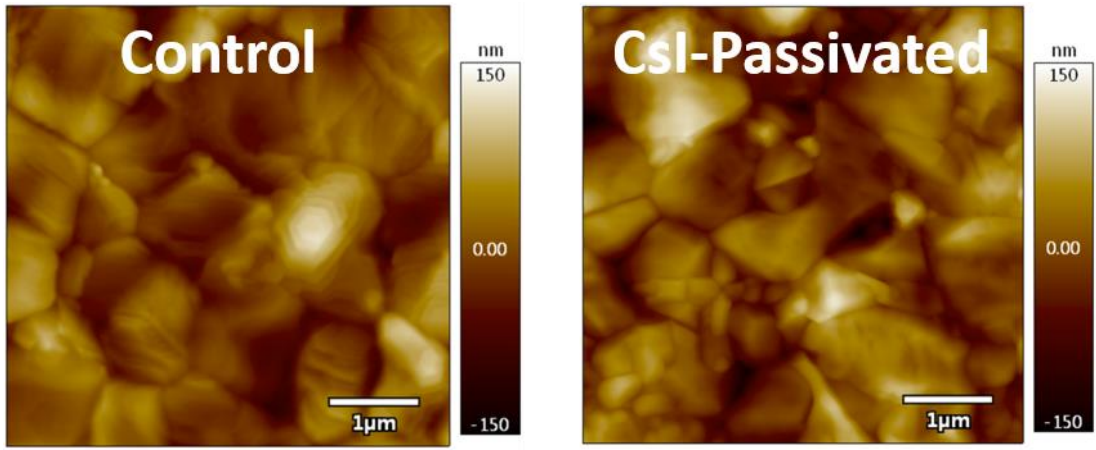

**Figure S12** AFM measurements of control and CsI-treated perovskite films exhibiting similar roughness

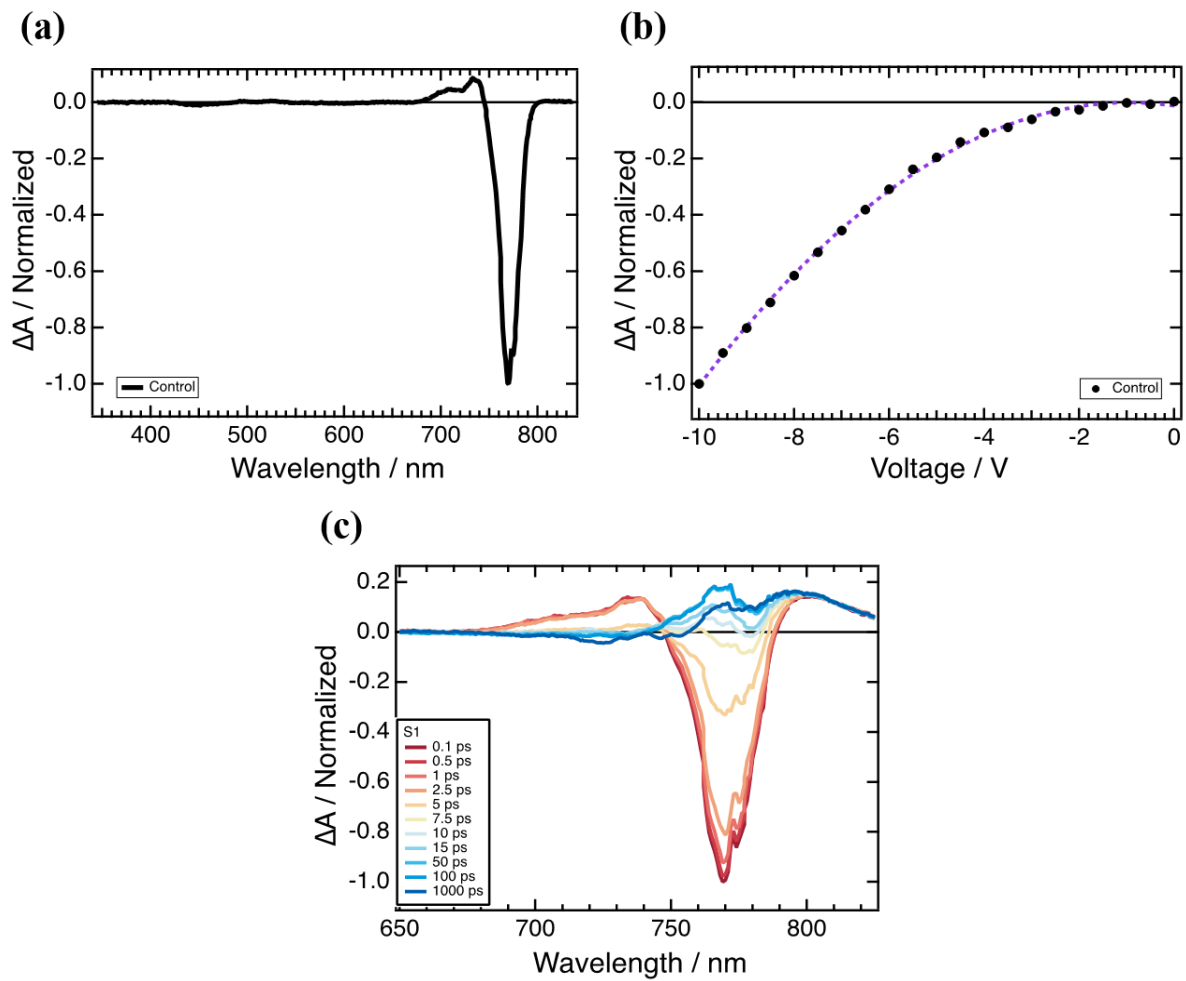

**Figure S13** (a) EA spectra detected at the second harmonic of the electric-field modulation frequency measured at 293 K (b) and at different voltages (c) TREA plots of the control perovskite film.

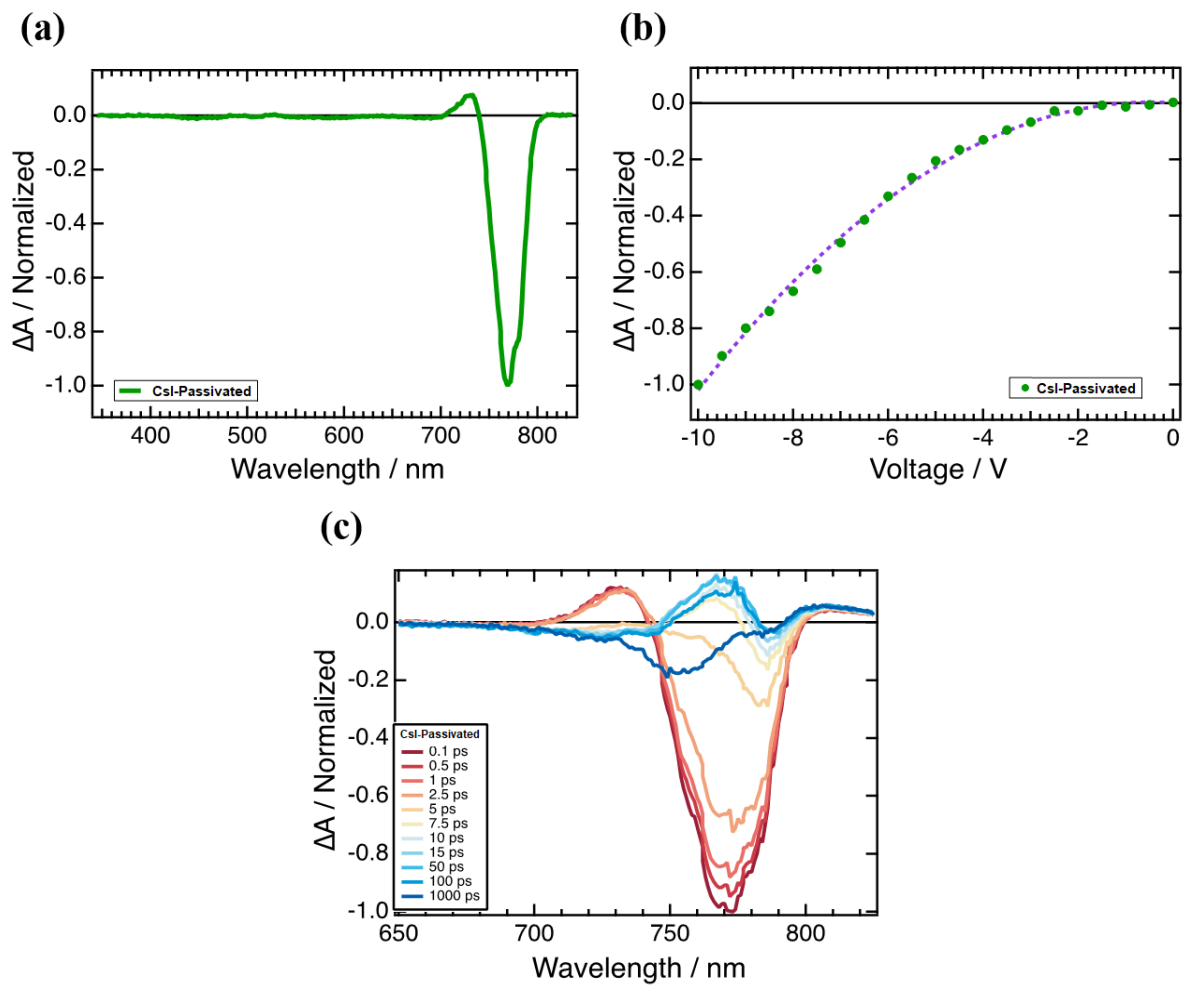

**Figure S14** (a) EA spectra detected at the second harmonic of the electric-field modulation frequency measured at 293 K (b) and at different voltages (c) TREA plots of the CsI-passivated perovskite film.